# CONSIDERATIONS FOR AUTOMATED MACHINE LEARNING IN CLINICAL METABOLIC PROFILING: ALTERED HOMOCYSTEINE PLASMA CONCENTRATION ASSOCIATED WITH METFORMIN EXPOSURE

Alena Orlenko, Ph.D.[*,1]; Jason H. Moore, Ph.D.[*,1]; Patryk Orzechowski, Ph.D.[1,2]; Randal S. Olson, Ph.D.[1]
*1 - Institute for Biomedical Informatics, University of Pennsylvania
Philadelphia, PA, USA;
2 - Department of Automatics and Biomedical Engineering, AGH University of Science and Technology,
Krakow, Poland*

Junmei Cairns, PhD[1]; Pedro J. Caraballo, MD[2]; Richard M. Weinshilboum, MD[1]; Liewei Wang, MD, PhD[1,†]
*1 - Department of Pharmacology and Experimental Therapeutics, Mayo Clinic
2 - Department of Endocrinology, Mayo Clinic
Rochester, MN, USA*
Email: Wang.Liewei at mayo dot edu

Matthew K. Breitenstein, Ph.D.[†]
*Institute for Biomedical Informatics and Center for Pharmacoepidemiology, Research, and Training,
University of Pennsylvania
Philadelphia, PA, USA*
Email: mkbreit at upenn dot edu

With the maturation of metabolomics science and proliferation of biobanks, clinical metabolic profiling is an increasingly opportunistic frontier for advancing translational clinical research. Automated Machine Learning (**AutoML**) approaches provide exciting opportunity to guide feature selection in agnostic metabolic profiling endeavors, where potentially thousands of independent data points must be evaluated. In previous research, AutoML using high-dimensional data of varying types has been demonstrably robust, outperforming traditional approaches. However, considerations for application in clinical metabolic profiling remain to be evaluated. Particularly, regarding the robustness of AutoML to identify and adjust for common clinical confounders. In this study, we present a focused case study regarding AutoML considerations for using the Tree-Based Optimization Tool (**TPOT**) in metabolic profiling of exposure to metformin in a biobank cohort. First, we propose a tandem rank-accuracy measure to guide agnostic feature selection and corresponding threshold determination in clinical metabolic profiling endeavors. Second, while AutoML, using default parameters, demonstrated potential to lack sensitivity to low-effect confounding clinical covariates, we demonstrated residual training and adjustment of metabolite features as an easily applicable approach to ensure AutoML adjustment for potential confounding characteristics. Finally, we present increased homocysteine with long-term exposure to metformin as a potentially novel, non-replicated metabolite association suggested by TPOT;

---

[*] Contributions equivalent for first authorship consideration
[†] Contributions equivalent for senior and corresponding authorship consideration

an association not identified in parallel clinical metabolic profiling endeavors. While warranting independent replication, our tandem rank-accuracy measure suggests homocysteine to be the metabolite feature with largest effect, and corresponding priority for further translational clinical research. Residual training and adjustment for a potential confounding effect by BMI only slightly modified the suggested association. Increased homocysteine is thought to be associated with vitamin B12 deficiency – evaluation for potential clinical relevance is suggested. While considerations for clinical metabolic profiling are recommended, including adjustment approaches for clinical confounders, AutoML presents an exciting tool to enhance clinical metabolic profiling and advance translational research endeavors.

## 1. Background

### 1.1. *Introduction to metabolomics and study motivation*

Metabolomics, the study of organic chemical signatures within a specimen, has been increasingly deployed in clinical research applications. Characterization of perturbations to the metabolome (a.k.a. phenome) hold great promise to elucidate novel biomedical insights and potential disease mechanisms. While many 'omics perspectives provide unique molecular insights, the phenome reflects biological perturbation closest to clinical phenotype manifestation. With the proliferation of biobanks [1] – where consenting patients voluntarily donate a wide-array of biologic specimens (e.g. blood, urine, saliva) to be systematically stored and utilized for research – opportunities for secondary research applications, including metabolomics, using primary specimens abound [2]. As both the science of metabolomics advances and scale of biobanks increase, clinical metabolic profiling holds increasing promise to identify novel biological insights regarding disease state, drug response, and clinical heterogeneity [3].

Metabolic profiling is a multi-step process that 1) initiates with analytical chemistry measurement {e.g. liquid chromatography (**LC**), mass spectrometry (**MS**), Nuclear magnetic resonance spectroscopy (**NMR**)}, including deployment of tandem techniques such as LC/MS, of organic compounds contained within a biological specimen; 2) algorithmic association of raw measurements with known discrete metabolites; 3) establishment of relative metabolite concentrations; and 4) concluding with statistical generation of a profile of metabolites (i.e. metabolic profile) perturbed within the phenome given an exposure of interest. Metabolic profiling of both disease [4] and drug exposures [5] have successfully identified distinct signatures. While some of these features have been shown to remain stable over time [6], some metabolites and physiologic states are known to rapidly fluctuate [7]. Further, some metabolites are known to be lipid soluble, with measured concentrations noticeable altered in patients with elevated BMI [8] – a biological rationale for potential confounding by BMI in clinical metabolic profiling. For agnostic untargeted metabolic profiling, thousands of metabolites are identified, whereas for targeted metabolic profiling, only a small group of metabolites are selected *a priori* based on hypothesized biological relevance. With untargeted metabolic profiling, distinct analytical challenges remain as thousands of potentially unique features are ascertained, frequently exceeding the number of samples analyzed. Augmenting the metabolic profile with other 'omics perspectives only further enhances this complexity. Regardless of selected approach and application, great opportunity exists for semi-automated machine learning approaches to assist in agnostic selection and inclusion of features in metabolic profiling endeavors. Our current application is focused within the later part of the targeted metabolic profiling process, characterizing long-term exposure to the drug metformin as a monotherapy within a human biobank cohort.

### 1.2. *Automated Machine Learning and TPOT*

Machine Learning (**ML**) approaches hold great opportunity to enhance metabolic profiling endeavors. Tree-based optimization tool (**TPOT**), our specific tool of interest, is an Automated Machine Learning (**AutoML**) tool with recent demonstrable success. Specifically, TPOT has been observed to automatically optimize ML pipelines that match or exceed the performance of traditional supervised approaches [9, 10, 11] while requiring minimal adjustments to default parameters. The mixture of data types deployed in human metabolic profiling and expansive feature space are ideally suited for enhancement with AutoML approaches. In genomics applications [9], TPOT has delivered promising predictive performance while being demonstrably robust to mixed datatypes with large feature spaces. Given the mixture and expansive feature space of data types found in clinical metabolic profiling, we posit that AutoML approaches offer opportunity for a robust, agnostic profiling solution. However, a thorough evaluation of potential caveats and considerations for application of AutoML using TPOT in clinical metabolic profiling is necessary. This includes specific considerations regarding continuous metabolite features in a potentially expansive feature space.

In this study, we provide an annotated methodological case study applying AutoML in clinical metabolic profiling of patients exposed to metformin monotherapy. Patient data was collected previously for traditional clinical metabolic profiling endeavors [12] from patients nested within a biobank cohort [2] Highlighted within our methodological case study are the following items: 1) A focused overview of clinical metabolic profiling using automated supervised machine learning methods. Specifically, demonstrated using the TPOT tool; 2) Necessary pre-processing and analysis steps for development of a clinical metabolic profile using AutoML; 3) Current state of the art for identification of confounding characteristics using AutoML; 4) Proposed strategies to adjust for confounding characteristics of different types commonly encountered in clinical metabolic profiling; 5) Finally, propose an AutoML-based tandem rank-accuracy metric for agnostic data-driven feature selection in clinical metabolic profiling.

## 2. Methods

All analyses and experiments, described in-depth below, were conducted using Python programming packages Sklearn, Pandas, and Numpy. All figures were generated using Python Matplotlib and

Seaborn programming packages. TPOT v0.8 software [9,10,11] <https://github.com/rhiever/tpot> was exclusively utilized for AutoML experiments.

## 2.1. *TPOT overview*

Fundamentally, TPOT takes a supervised learning dataset as input and recommends a series of preprocessing, feature construction, feature selection, and ML modeling operations that maximize the predictive performance of the final ML model. We call this series of operations a pipeline. TPOT optimizes the analysis pipeline using a stochastic optimization process that begins with several simple, random pipelines (the *population*). For every iteration of the optimization process (a *generation*), TPOT makes several copies of the current best-performing pipelines in the population and then applies random changes to them, such as adding or removing an operation or tuning a parameter setting of one of the operations. These stochastic changes can have positive or negative effects on the performance of the pipelines, and as such allow TPOT to explore new analysis pipelines that were never previously considered. At the end of every generation, the worst performing pipelines are removed from the population and TPOT proceeds to the next generation. After a fixed number of generations (in this study, 1,000 generations), TPOT recommends the best-performing pipeline that it ever created during the optimization process. In this study, we present observations from the TPOT pipeline as described. For more details regarding the TPOT algorithm and tool, see [9,10,11] and the software package online at https://github.com/rhiever/tpot

## 2.2. *TPOT default parameters and pre-processing*

A series of pre-processing steps were initiated to evaluate sensitivity of TPOT for detection of common clinical confounders (e.g. age, gender, body mass index (**BMI**), batch effects). Supervised classification analysis was performed using an out-of-the-box TPOT deployment. A classification predictive model was generated on the full dataset containing both prioritized metabolite features and clinical covariates using the following settings: number of generation 2000, population size 1000, 5-fold cross validation on the training set, and standard accuracy as a performance metrics. Prior to TPOT analyses, the cohort was randomly stratified into separate 75% training and 25% testing datasets. A unique random seed was selected for each of the 5 independent replications. Resulting models were characterized using accuracy metrics, representing the fraction of corrected prediction with the best possible score 1.0. For each replicate, feature importance was measured and rank was assigned in accordance with importance coefficients. Ranks were summed across replicates where inverse of sum of ranks served as the metric for feature importance across experiments. Specifically, rank coefficient or $r_x$, where x is a feature coefficient from replicate *i*, *n* – total number of the TPOT replicates: $r_x = 1/ \sum_{i=1}^{n} x$

### 2.3. *TPOT Analysis*

TPOT models generated predictive ranks, an approximation of relative effect size generalizable across TPOT-selected machine learning algorithms, for comparing importance of individual features. Model performance overall was evaluated using $R^2$, or coefficient of determination (i.e. accuracy), describing the fraction of response variance described by the model, with a maximum possible score of 1.0. In our work, we highlighted the potential utility of rank-accuracy measures deployed in tandem to guide agnostic feature selection in clinical metabolic profiling.

To evaluate TPOT's automatic adjustment capabilities in clinical metabolic profiling, we evaluated the following features for potential confounding: 1) BMI – metabolites evaluated using case-only (metformin monotherapy exposed) and stratified {split by the median value of BMI (2 groups) and common clinical thresholds (<18.5, 18.5-25, 25-30, ≥30)} datasets; 2) Batch effect – 8 splits were applied using TPOT classification mode with case status set as the target variable. For each replicate, feature importance was measured and predictive ranks were assigned; 3) Dose-dependent metabolite effect of metformin exposure – case-only analysis was performed where prescribed metformin daily dose and measured metformin plasma concentration were included; 4) Confounding associations (i.e. sensitivity) not identified by TPOT (i.e. insensitive) – metabolites were adjusted using either stratification or residual adjustment approaches, both of which are described in-depth below.

To demonstrate the utility of TPOT for feature selection towards ascertainment of a metabolic profile, classification TPOT analysis was then applied to a reduced-feature dataset containing only the prioritized metabolite features (i.e. potential confounding clinical covariates were removed). TPOT settings described in the previous section were utilized. Predictive ranks were deployed to aid in feature selection of metabolic profiles; metabolites were sorted in accordance to their rank coefficient and recursive feature elimination estimated the strength (e.g. accuracy score) of prediction for various consecutive combination of sorted features. Pipelines from TPOT analyses evaluated performance of various feature sets by reporting training and testing set accuracy. To understand the impact of potential confounding insensitive to TPOT, the ascertained clinical metabolite profile was replicated using a BMI-adjusted dataset, where metabolite measured concentrations were replaced with residuals from independent univariate linear regression models of individual metabolites and BMI.

## 3. Results

### 3.1. *Cohort characteristics*

Exposure to clinically stable (i.e. 'long-term') metformin monotherapy was profiled using a de-identified case-control dataset representing 546 unique patients nested within a biobank cohort. All data was previously collected for parallel metabolic profiling endeavors [12]; data was de-identified by our collaborating institution prior to release for analysis. IRB coverage for both prior research data collection (original study purpose) [Mayo Clinic 15-003347 and 08-007049] and secondary analysis [Penn 827996] were obtained. A pre-selected panel of amine-based metabolites (n=42) were previously measured from human plasma samples using tandem LC-MS. Clinical features were previously ascertained using electronic health record(**EHR**)-based phenotyping and confirmed by manual chart review. Clinical features included common covariates (age, gender, BMI, and metabolite batch) and metformin exposure (metformin prescribed daily dose and metformin plasma concentration). Cases (n=273) included patients exposed to metformin monotherapy with type 2 diabetes having glycemic control; controls consisted of healthy normal patients with no known metformin exposure. Case and control patients were previously matched by age and gender prior to sample selection and were statistically randomized and assigned to batches prior to metabolomic measurement.

### 3.2. *Descriptive analyses using default TPOT parameters*

Univariate Pearson's correlations were generated for metformin exposure (case-control status) using TPOT (**Figure 1**). Metformin exposure demonstrated correlations with varying effect and direction for both metabolite (e.g. alanine and citrulline) and clinical (BMI, metformin prescribed daily dose, metformin plasma concentration) features. These results demonstrated that associations exist within the dataset, suggestive of potential confounding between metformin exposure and metabolic perturbation. Further, strong associations were identified for several metabolite-metabolite pairings. For example, tyrosine, valine, isoleucine, leucine and phenylalanine were positively correlated ($r > 0.5$) with each other. While physiologically unclear, such distinct clustering of associations might suggest potential for proximal, interrelated metabolic responses.

From the perspective of rank associations, we evaluated sensitivity of TPOT to identify potential confounding features using default parameters. Not surprisingly, increased prescribed daily dose of metformin was associated with increased relative effect (~ 5 times larger than the most predictive metabolite) for predicting metformin exposure (**Figure 2A**). Since prescribed metformin daily dose was a non-binary feature, characterized by 10 discrete values (ranging from 250 to 3000 mg), adjusting for a potential confounding effect using stratification alone – one of the common strategies to adjust for confounding feature in AutoML analysis – had potential to create unbalanced or underpowered subgroups and bias resulting associations. We posited, while increased dose has potential to mask identification of relevant metabolite features, inherently enhancing the biological effect of metformin exposure, it is unlikely to introduce bias by a confounding effect and can be removed from analysis. When dose and concentration of metformin exposure were removed from consideration, a more gradual distribution of rank coefficients were observed. The top three features contained the metabolites homocysteine and citrulline and clinical feature BMI (**Figure 2B**). These findings, together with existing biomedical knowledge, suggest that dose-dependent features have potential to mask important metabolite features and BMI might introduce bias due to confounding.

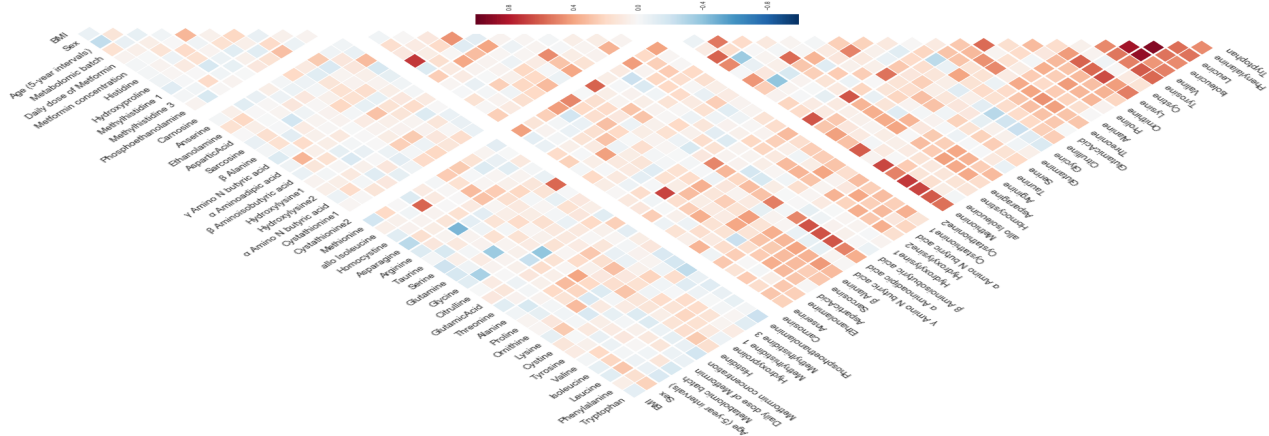

**Figure 1**. Pearson's correlation coefficients for metabolite and clinical features.

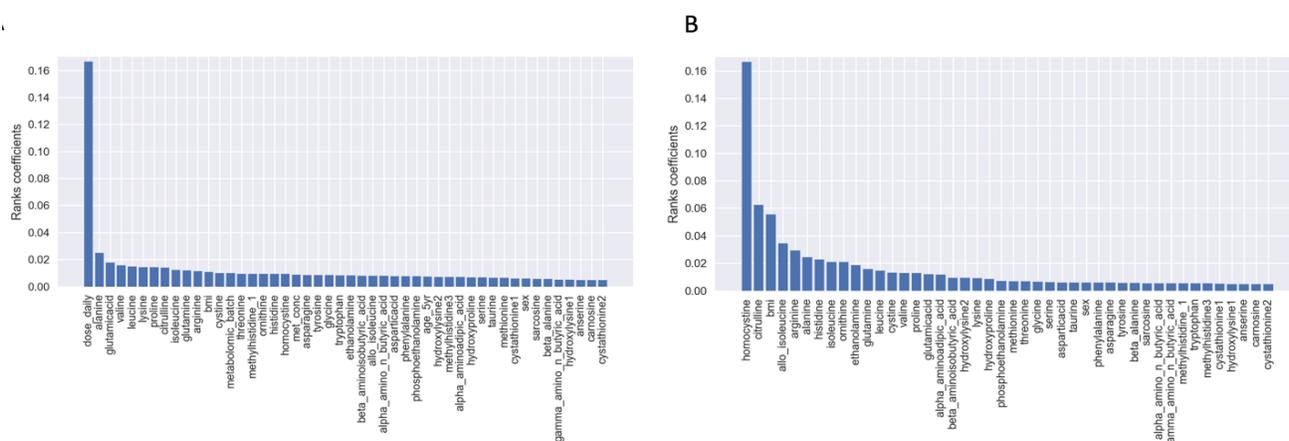

**Figure 2**. Metabolite and clinical feature ranks. The most predictive features have lowest values; the least predictive features have highest values. A) All metabolic and clinical features. B) All metabolic and clinical features excluding daily dosage.

### 3.3. *Evaluation of clinical characteristics for potential confounding*

Within the below sections, we assessed potential confounding using out-of-the-box AutoML. Specifically, we evaluated BMI, metabolomics batch effect, and potential dose-dependent effects:

**3.3.1.** *Body mass index (BMI).* While BMI was demonstrated to be associated with metformin exposure overall, suggesting confounding, potential for within case-control status confounding was unknown. To elucidate potential within-case confounding (i.e. confounding by indication) by BMI, TPOT regression analysis was performed on a case-only dataset with metabolite and BMI features and evaluated using $R^2$ or accuracy (**AC**) metrics (**$R^2$= testing; training**). TPOT generated various regression models, including Elastic Net Regressor with built-in Cross-Validation, Extra Tree Regressor, Random Forest Regressor, and Ridge Regressor with built-in Cross-Validation. Due to low accuracy, the highest being ($R^2$=0.48;<0.01), we were unable to assign rank coefficients or an effect. This low accuracy suggested that either the TPOT models were insensitive to within-case BMI or that no confounding effect existed. To further elucidate a potential effect, BMI was evaluated with various splits and thresholds. However, the highest accuracy ($R^2$= 0.89;neg) implicated a model likely over-fit and containing false positives. Univariate linear regression analysis was performed on the same datasets to serve as a benchmark for TPOT regression performance. In independent regression analyses of top metabolite features, model performance remained poor for alanine ($R^2$=0.58;null) and α-aminoadipic acid ($R^2$=0.62;0.42). While the models did not identify BMI associations, the existence of distinct distributions of BMI within cases and controls (**Figure 3**) suggest potential confounding and potential insensitivity of TPOT where collinearity exists.

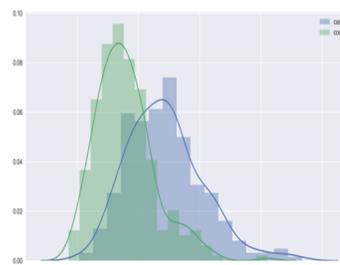

**Figure 3.** Distribution of BMI distribution within case and control

**3.3.2.** *Metabolomics batch effect.* To elucidate a potential batch effect, TPOT analysis was performed stratified by batches. We ran TPOT classification analysis for each batch subset and compared performance (**Figure S1**). Overall, subsets performed very well with high accuracy for both training and testing sets, suggesting strong potential for a batch effect. For individual batch performance, we observed the following: batch 1(AC=0.90;0.87); batch 2 (AC=0.90;0.82); batch 3 (AC=0.95;0.94); batch 4 (AC=0.95;0.81); batch 5 (AC=0.92;0.90); batch 6 (AC=0.94;0.93); batch 7 (AC=0.96;0.73); batch 8 (AC=1.0;1.0). However, case-control frequencies varied within these associations, with batch 2 having 61 cases and batch 5 having only 2 cases. Unbalanced randomization between case-control selection in batch assignment likely contributed to a potential batch effect.

**3.3.3.** *Metformin dose-dependent effect.* Case-only TPOT analysis generated strong dose-dependent associations (prescribed metformin daily dose and measured plasma metformin concentrations) across several TPOT-generated models. However, when benchmarked to univariate associations, both training and testing accuracies were very low (AC<0.11), suggesting likely model overfitting. In context, these findings suggest that while dose effects may mask associations in clinical metabolic profiling due to being contained within the exposure, the observed is potentially not a true confounding effect. Further work remains to robustly identify and adjust for dose-dependent effects in AutoML.

### 3.4. *Obtaining metabolic profile guided by predictive ranks and tandem-rank accuracy*

In development of a clinical metabolic profile of metformin exposure, TPOT models selected Gradient Boosting Classifier and Extra Trees Classifier for classification task with accuracy scores for a training set 0.98 and above and for a testing set 0.83 and above. Distribution of rank coefficients clearly prioritized homocysteine and citrulline as top metabolite features. To ascertain additional metabolite features of potential relevance, recursive feature elimination was applied and features sorted in accordance by their rank. Training and testing accuracy of the model continued to increase (**Figure 4**) up to the addition of the top 4 important metabolite features (homocysteine, citrulline, allo- isoleucine, and arginine) and remained relatively unchanged beyond inclusion of an additional 2 metabolite features (alanine and isoleucine).

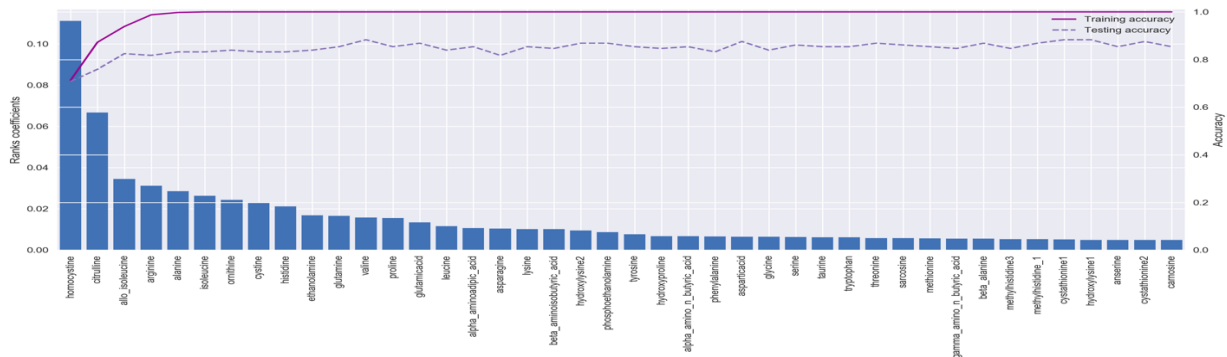

**Figure 3**. AutoML generated clinical metabolic profile for exposure to metformin guided by tandem-rank accuracy measure. Sorted histogram of predictive power for metabolite inverse (for ease of interpretation) sum of ranks (blue bar), training set accuracy, (solid magenta line), and testing set accuracy (dashed magenta line) describe relative feature effect size and model performance.

In this analysis, the rank metrics provided information about relative importance of metabolic variables with respect to their predictability of the outcome (metformin exposure) variable. The accuracy metrics provided an estimate of the model performance on the various subsets of the features and help to distinguish the most discriminative features in the dataset. Together rank and accuracy generated a statistical support for distinguishing metabolites that show differential response to exposure. In this study, the tandem metrics demonstrated that homocysteine was consistently identified as our top TPOT-recommended feature, with a much larger magnitude of effect than other features.

### 3.5. *Proposed adjustments for confounding bias in AutoML analyses*

In the previous section, we identified features with potential to bias AutoML-based clinical metabolic profiling endeavors due to confounding effects. We identified that TPOT is potentially insensitive to identification of low-effect confounding features, and that high-effect features may mask potentially relevant prioritized metabolite features. As such, manual adjustment for potential confounding features might be necessary. Further, as is often required for epidemiological inquiry, select feature adjustment might be required to rule out a suspected confounding effect.

To adjust for confounding in AutoML, we suggest two data type-dependent adjustment strategies: 1) For continuous values, we propose that residuals obtained from independent linear regressions [13] (e.g. between metabolites and BMI) be obtained prior to AutoML analyses. This aims to be consistent with approaches appropriately address confounding in multivariate statistical analyses [14]. In our application, the residual index was computed independently for each metabolite as the residual from the simple linear regression of metabolite variable on the confounding variable (i.e. BMI). The residual distances of individual points from regression line then served as the estimators of metabolites. Our adjustment for BMI slightly enhanced our homocysteine association (**Figure S2**) – demonstrated to be increased in plasma concentration with metformin exposure in multivariate linear regression – with the model accuracy remaining comparable. Metabolite features originally ranked below homocysteine in unadjusted analysis consistently remained below homocysteine, but were slightly modified by rank order and magnitude. We recommend sensitivity analysis and regression diagnostic methods to select a proper regression model for adjustment in AutoML applications. 2) For categorical data types, we recommend stratification, where independent analyses are conducted and findings (e.g. means) presented in aggregate. In this approach, AutoML generates feature importance coefficients for each subset and then transformed into ranked coefficients (described in Section 2.2) where mean of ranks are calculated over all subsets. However, stratification is known to be negatively impacted by low sample power. Frequently, well-powered strata produce more accurate estimates than relatively lower powered strata. To supplement this deficiency, we suggest weighted mean, particularly Mantel-Haenszel [15], where strata are prioritized be statistical power. When applied, stratum-specific adjusted relative risk estimates can be calculated, providing an overall summary measure of effect. Approach described for categorical data types might be most aptly applied in future research to elucidate potential metformin dose-dependent effects.

## 4. Discussion

In this study, we demonstrated AutoML considerations using TPOT for metabolic profiling of exposure to metformin monotherapy in a biobank cohort. Our two major informatics contributions include: 1) tandem rank-accuracy measures to guide agnostic feature selection and corresponding threshold determination in clinical metabolic profiling endeavors, and 2) residual training and adjustment of metabolite features in AutoML analysis. Both our informatics contributions and identified metabolite associations contribute to precision medicine knowledge.

### 4.1. *Considerations and adjustments for confounding features*

In our analysis, we demonstrated that while AutoML is a potential powerful tool for clinical metabolic profiling, specific considerations and adjustments might need to be applied for potentially confounding characteristics. Correlation and agnostic TPOT analyses demonstrated that daily dose and BMI had strong-to-medium associations with metformin exposure. While not a focus of our analysis, this dataset is uniquely suited for future evaluation of dose-dependent effects using AutoML and other analysis approaches. For datasets with potential confounding, we proposed two data type-dependent adjustment strategies: 1) stratification for categorical features, and 2) independent residual identification and application for continuous features.

### 4.2. *Future study design considerations*

AutoML methods (and ML methods in general) can be sensitive to the dataset quality in terms of sample size and sample structure. A distinct strength of our study was the well-powered (n = 546) dataset with targeted metabolites. Conversely, datasets with a small sample size (n < 50 samples) can often lead to overfitting, especially when the dataset has high variance due to random noise. Here we had 546 samples and 42 metabolites, which is considered a good ratio of features to samples to avoid high variance problem. However, in untargeted clinical metabolic profiling studies, where the number of metabolites exceeds the number of samples 10 or even 100-fold, this high variance is a common problem. In this scenario, even high accuracy scores could be unreliable without deploying an alternative strategy. One suggested approach to avoid this pitfall is to apply feature selection methods before running AutoML analysis. Relief-based algorithms, recursive feature elimination and regularization techniques are among the most common approaches used to treat overfitting and is directly applicable in future untargeted clinical metabolic profiling endeavors.

Imbalanced datasets with one phenotype overrepresented (i.e. more controls than cases) can also cause bias in AutoML and ML classification tasks. Several adjustments could be made including: 1) changing performance metrics to one that can give more insight into the accuracy of the model than traditional classification accuracy (e.g., area under the ROC curve, precision and recall); 2) resampling data using either addition of copies of samples from the under-represented phenotype or removal of samples from the over-represented phenotype. Imbalanced datasets can also make it difficult to apply stratification approaches to control for confounding. A potential weakness of our study is that batch effect could not be reasonably incorporated into adjustment approaches due to imbalance. Ensuring balanced observation batches is a critical consideration for study design in future clinical metabolic profiling studies.

Finally, datasets with a large percentage of missing values can be problematic for AutoML and ML methods. Another strength of our study includes that all features were fully populated. Many ML methods cannot handle missing values by default, so a common approach is to replace all missing values in a column with the median or mean value of that column (for all columns with missing values), or even replace all missing values with a fixed value (e.g., -99 or 0, depending on the feature). However, replacing missing values in such a manner, especially when there is a large percentage of missing values can introduce noise into the dataset and bias analysis. Thus, we recommend taking a thorough approach when replacing missing values in a dataset for AutoML and ML.

### 4.3. *Increased homocysteine*

Beyond our specific informatics contributions, we demonstrated the utility of AutoML to enhance multi-omic perspectives in pursuit of precision medicine knowledge. In this study, we present increased homocysteine with long-term exposure to metformin as a potentially novel metabolite association suggested by TPOT; an association not identified in parallel clinical metabolic profiling endeavors. While warranting independent replication, our tandem rank-accuracy measure suggests homocysteine to be the metabolite feature with largest effect, and corresponding priority for further translational clinical research. Residual training and adjustment for a potential confounding effect by BMI only slightly modified our initial association. Elevated homocysteine levels are clinically associated with vitamin B12 and folate deficiency [16] – we suggest future consideration for potential clinical relevance and independent replication. Elevated homocysteine is also associated with some increased posited risk for atherosclerotic disease [17], potentially cancer [18], and depression [19], but has insufficient evidence to suggest consideration as a clinical predictor or biomarker.

While indeed, the maturation of metabolomics science and proliferation of biobanks are exciting, combining with the expansive clinical perspectives offered by EHR linkages offer unprecedented opportunity. We posit that EHR perspectives of phenotypic divergence combined with metabolic variation are poised to become powerful facets advancing clinical translational science. Judicious application of ML and AutoML approaches will become increasingly powerful in multi-omic research.

## 5. Conclusion

AutoML is an exciting tool holding great promise to enhance clinical metabolic profiling and advance translational research endeavors; considerations are recommended, including adjustment approaches for clinical confounders. Our identified association of increased homocysteine with long-term metformin exposure warrants independent replication and evaluation for potential clinical relevance.

**Acknowledgments.** This research was made possible with generous support from the Mayo Clinic, Center for Individualized Medicine. The Institute for Biomedical Informatics, University of Pennsylvania and Mayo Clinic Cancer Genetic Epidemiology Training Program (R25 CA092049) further supported this research.

## 6. Supplementary Information

### 6.S1. *Batch effect analyses*

Supplementary figures regrading Section **3.3.2. *Metabolomics batch effect***: Histograms of metabolite features predictive ranks for batch-stratified subsets (Batch 1-7) generated by AutoML. TPOT analysis was performed stratified for batch-stratified subsets to elucidate a potential batch effect and model performance evaluated. For individual batch performance (**Supplementary Figures S1.1-S1.7**), we observed the following: batch 1(AC=0.90;0.87); batch 2 (AC=0.90;0.82); batch 3 (AC=0.95;0.94); batch 4 (AC=0.95;0.81); batch 5 (AC=0.92;0.90); batch 6 (AC=0.94;0.93); batch 7 (AC=0.96;0.73). However, case-control frequencies varied within these associations dramatically (i.e. batch 2 having 61 cases and batch 5 having only 2 cases).

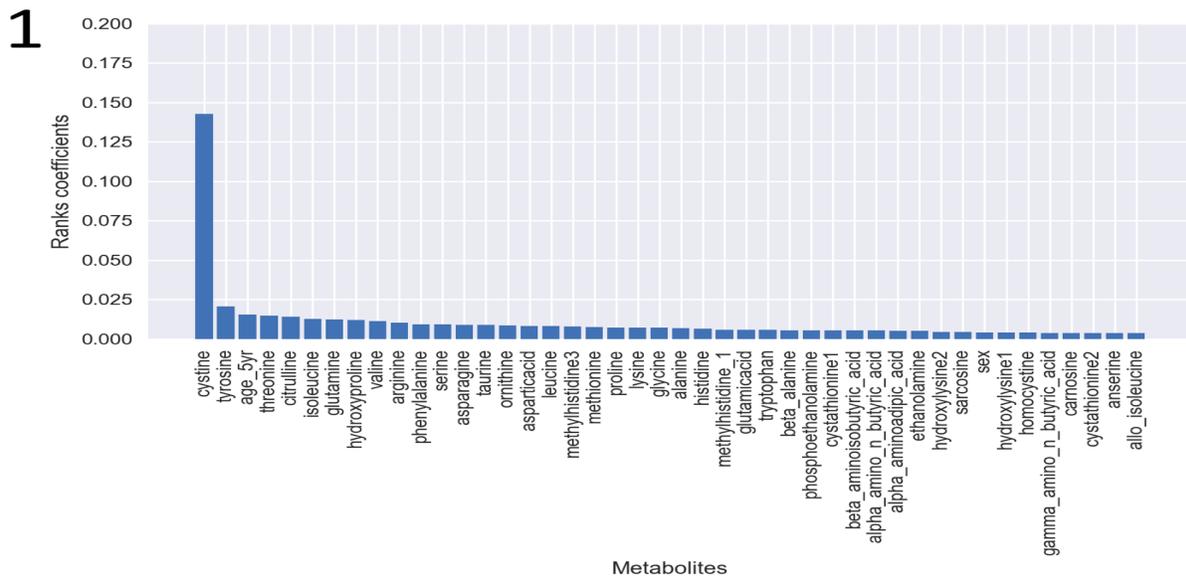

**S1.1**. Metabolite and clinical feature ranks generated by TPOT for batch 1.

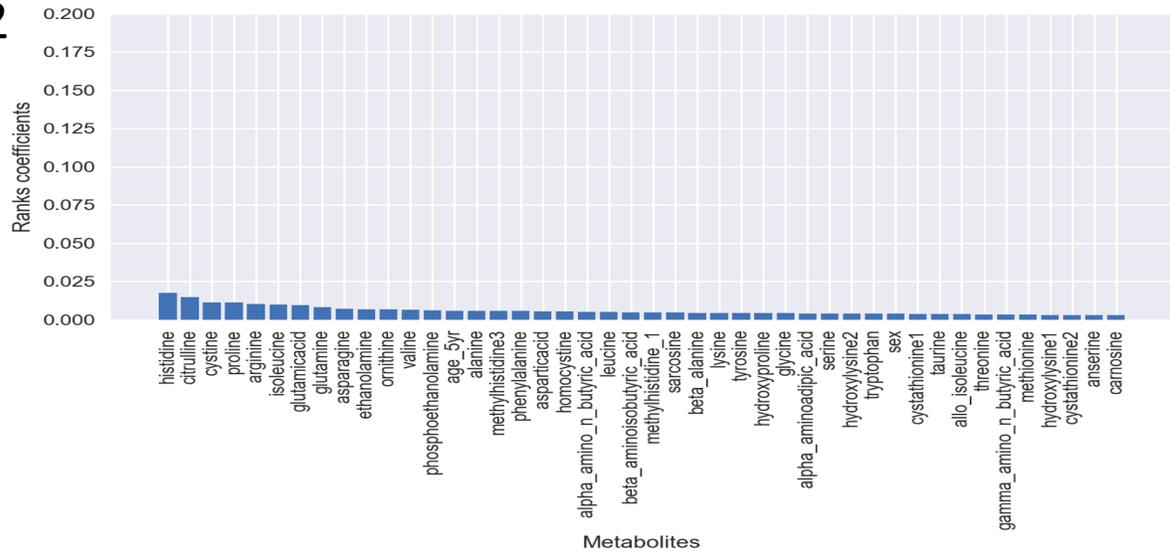

**S1.2**. Metabolite and clinical feature ranks generated by TPOT for batch 2.

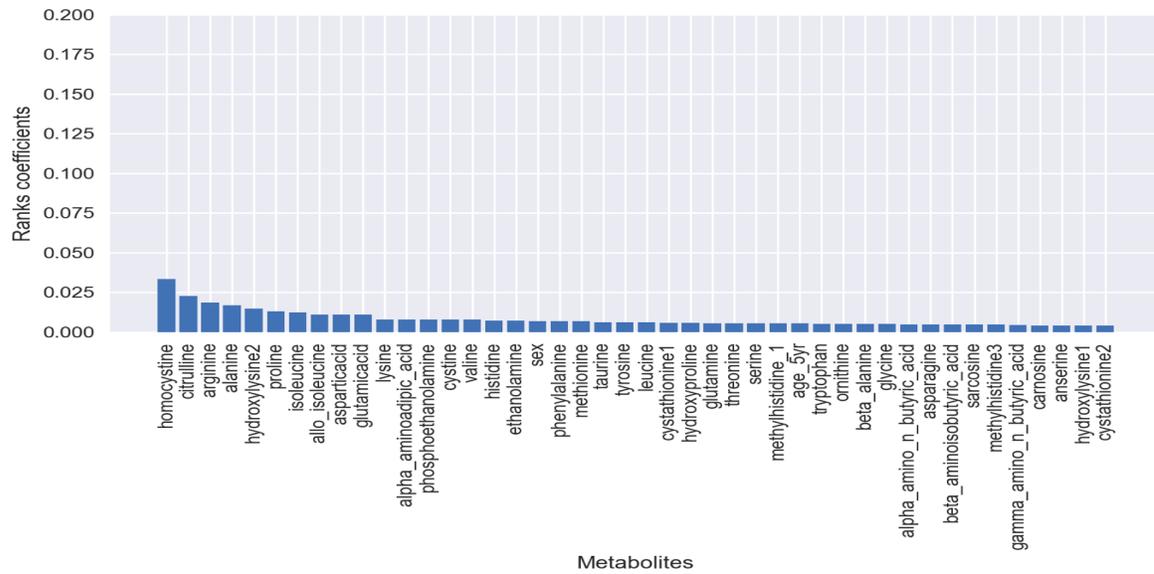

**S1.3**. Metabolite and clinical feature ranks generated by TPOT for batch 3.

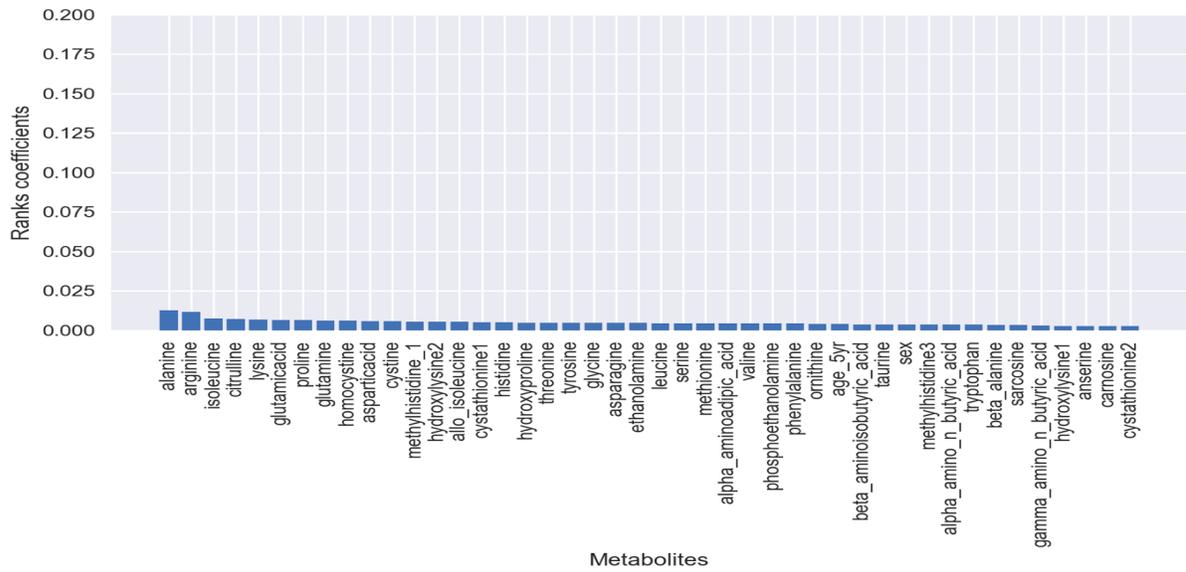

**S1.4**. Metabolite and clinical feature ranks generated by TPOT for batch 4.

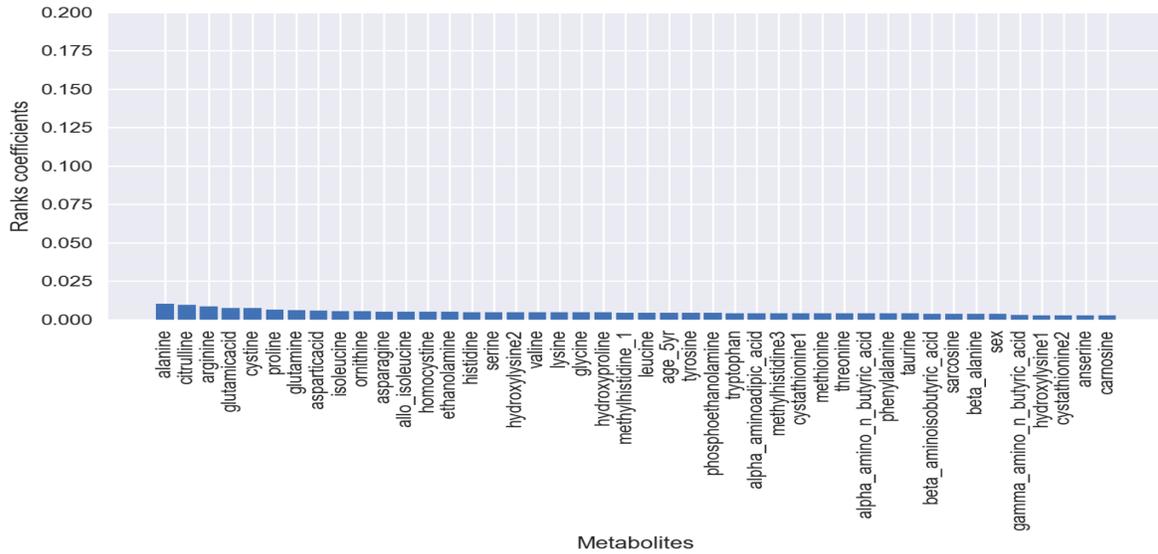

**S1.5**. Metabolite and clinical feature ranks generated by TPOT for batch 5.

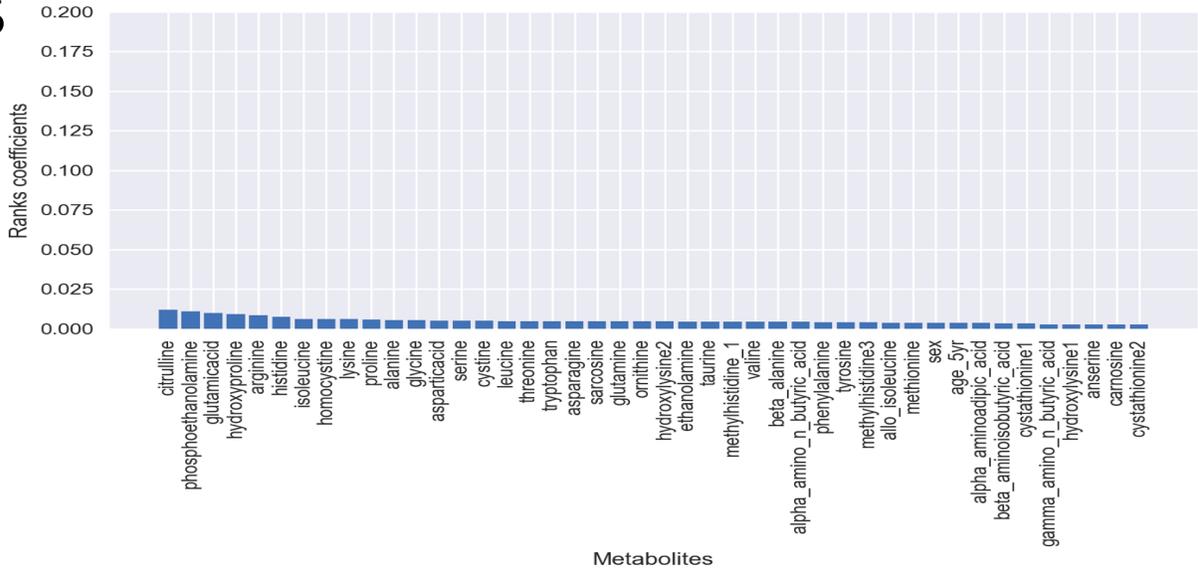

**S1.6**. Metabolite and clinical feature ranks generated by TPOT for batch 6.

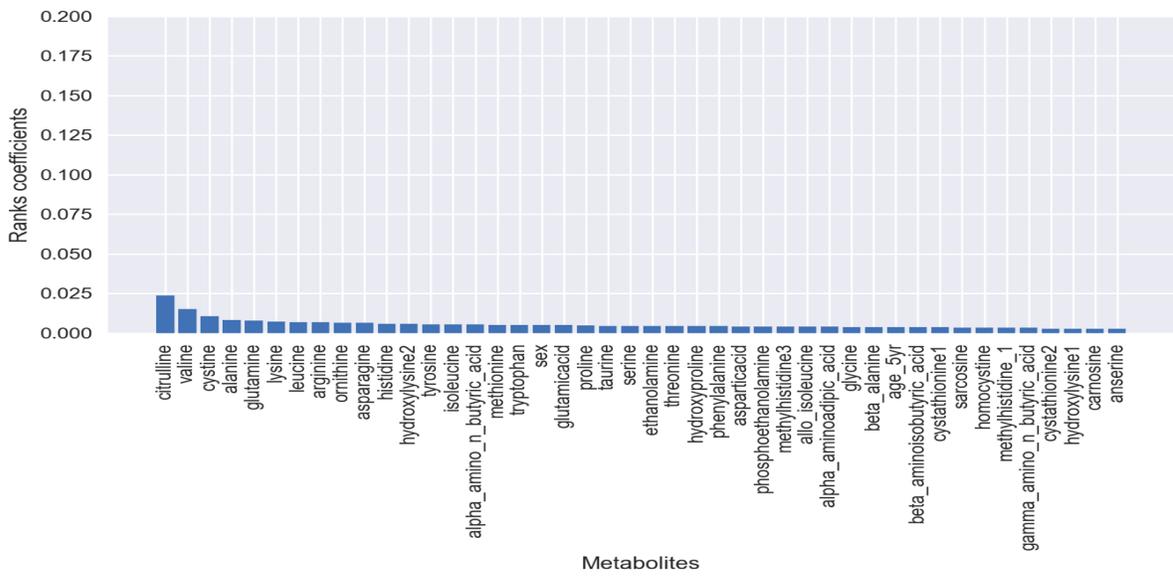

**S1.7**. Metabolite and clinical feature ranks generated by TPOT for batch 7

## 6.S2. *BMI-adjusted metformin clinical metabolic profile, tandem-rank accuracy measures*

Leading into **Section 3.5**, our analyses were unable to disambiguate if potential confounding by BMI did not exist within case (metformin monotherapy exposed) and control (no metformin exposure) status – or – if TPOT lacked sensitivity to automatically adjust for BMI. Demonstrating our proposed manual adjustment strategies for confounding in AutoML clinical metabolic profiling, posited to be robust given either scenario, we performed the proposed *residual adjustment* for BMI using linear regression, where true measures of metabolite concentrations were replaced with residuals from independent univariate linear regression models of individual metabolites (independent variable) predicting BMI (dependent variable). A second dataset containing independent BMI-adjusted metabolite proxy measures were generated. A BMI-adjusted clinical metabolic profile was then generated using the BMI-adjusted dataset (**Figure S.2.**). Our BMI adjustment only slightly enhanced the homocysteine signal – increased plasma concentration with metformin monotherapy exposure was identified without adjustment for BMI (**Figure 3**) – Overall model accuracy remained comparable between **Figure 3** and **Supplementary Figure S2**. However, metabolite features ranked below homocysteine were slightly modified by rank order and magnitude between adjusted and unadjusted analyses.

Our analysis suggests that true confounding by BMI (a form of confounding by indication) potentially did not exist within case (metformin monotherapy exposed) and control (no metformin exposure) status. Further, potential sensitivity of out-of-the-box TPOT to automatically adjust for confounding characteristics remains unclear; further evaluation in external datasets is needed to clarify. Residual adjustment (using linear regression) offers a viable solution to adjust covariates or assess potential confounding in agnostic clinical metabolic profiling.

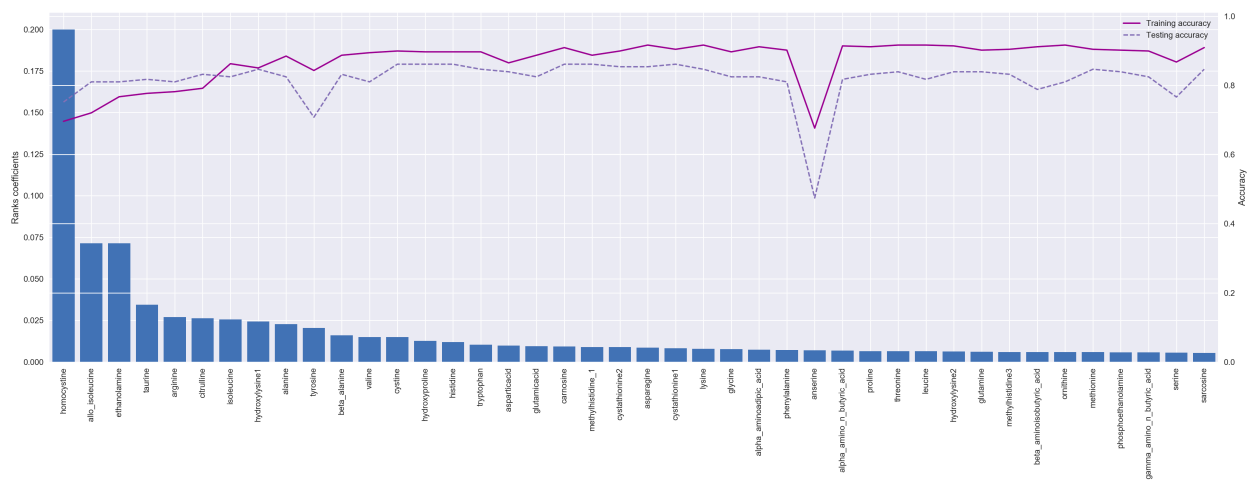

**S2**. **BMI-adjusted, AutoML clinical metabolic profile for exposure to metformin guided by tandem-rank accuracy measure**. Sorted histogram of predictive power for metabolite inverse (for ease of interpretation) sum of ranks (blue bar), training set accuracy, (solid magenta line), and testing set accuracy (dashed magenta line) describe relative feature effect size and model performance.

**6.S3.  *Reproducibility***

**6.S3.1.  *Python scripts.*** To ensure reproducibility of analysis steps, we have up uploaded our Python scripts to the Breitenstein Lab GitHub page developed for the above TPOT-based clinical metabolic profiling:
https://github.com/BreitensteinLab/AutoMLMetabolicProfiling_PSB2018.

Python scripts were authored by Alena Orlenko, Ph.D.:
https://gist.github.com/desmidium

TPOT [9,10,11] v0.8 software was utilized exclusively for *in silico* AutoML experiments:
https://github.com/rhiever/tpot

**6.S3.2.  *Data access and support.*** Benefactor support through the Mayo Clinic Center for Individualized Medicine provided resources for primary data collection from biological specimens within the Mayo Clinic Biobank. National Institutes of Health funds partially supported primary and secondary analyses. External access to the metformin metabolomics datasets is proposed through COnsortium of METabolomics Studies (**COMETS**), an extramural-intramural partnership that promotes collaboration among prospective cohort studies that follow participants for a range of outcomes and perform metabolomic profiling of individuals.